\documentclass[iop]{emulateapj}

\usepackage{epstopdf}
\usepackage{graphicx}

\def \aap{Astron.~Astrophys.}
\def \apjl{Astrophys.~J.}

\def \apj{Astrophys.~J.}

\def \mnras{Mon.~Not.~Roy.~Astron.~Soc.}

\def \nat{Nature}

\newcommand{\lsim}{\,\rlap{\raise 0.35ex\hbox{$<$}}{\lower 0.7ex\hbox{$\sim$}}\,}
\newcommand{\gsim}{\,\rlap{\raise 0.35ex\hbox{$>$}}{\lower 0.7ex\hbox{$\sim$}}\,}

\def\la{\mathrel{\hbox{\rlap{\hbox{\lower4pt\hbox{$\sim$}}}\hbox{$<$}}}}
\def\ga{\mathrel{\hbox{\rlap{\hbox{\lower4pt\hbox{$\sim$}}}\hbox{$>$}}}}

\def\kms {\hbox{km\,s$^{-1}$}}

\newcommand{\wl}{$\lambda$}

\newcommand{\wll}{$\lambda \lambda$}

\def\Mo  {\hbox{$~\rm M_{\odot}$}}
\def\ccm {{\rm cm}^{-3}}

\begin{document}


\title{The Destruction of the Circumstellar Ring of SN 1987A}

\author{
Claes Fransson\altaffilmark{1}, Josefin Larsson\altaffilmark{2}, Katia Migotto \altaffilmark{1}, Dominic Pesce\altaffilmark{3}, Peter Challis\altaffilmark{4}, Roger A. Chevalier\altaffilmark{3}, Kevin France\altaffilmark{5,6}, Robert P. Kirshner\altaffilmark{4}, Bruno Leibundgut\altaffilmark{7}, Peter Lundqvist\altaffilmark{1}, Richard McCray\altaffilmark{8}, Jason Spyromilio\altaffilmark{7}, Francesco Taddia\altaffilmark{1},
Anders Jerkstrand\altaffilmark{9}, Seppo Mattila\altaffilmark{10,11}, Nathan Smith\altaffilmark{12}, Jesper Sollerman\altaffilmark{1}, J. Craig Wheeler\altaffilmark{13}, Arlin Crotts\altaffilmark{14}, 
Peter Garnavich\altaffilmark{15}, Kevin Heng\altaffilmark{16}, Stephen S. Lawrence\altaffilmark{17}, Nino Panagia\altaffilmark{18,19,20}, Chun S. J. Pun\altaffilmark{21}, George Sonneborn\altaffilmark{22} and Ben Sugerman\altaffilmark{23}
}


\altaffiltext{1}{Department of Astronomy, The Oskar Klein Centre,
              Stockholm University, Alba Nova University Centre, SE-106 91 Stockholm, Sweden  
              }
   \altaffiltext{2} {KTH, Department of Physics, and the Oskar Klein
Centre, AlbaNova, SE-106 91 Stockholm, Sweden}
 \altaffiltext{3} {Department of Astronomy, University of Virginia, P.O. Box 400325, Charlottesville, VA 22904-4325, USA}
   \altaffiltext{3} { ESO, Karl-Schwarzschild-Strasse 2, 85748 Garching, Germany}
   \altaffiltext{4} {Harvard-Smithsonian Center for Astrophysics, 60 Garden Street, MS-19, Cambridge, MA 02138, USA. }
\altaffiltext{5}  {Laboratory for Atmospheric and Space Physics, University of Colorado, 392 UCB, Boulder, CO 80309, USA}
\altaffiltext{6}  { Center for Astrophysics and Space Astronomy, University of Colorado, 389 UCB, Boulder, CO 80309, USA}
\altaffiltext{7}  { European Southern Observatory, Karl-Schwarzschild-Strasse, 2, D-85748 Garching bei MŸnchen, Germany}
\altaffiltext{8}  { Department of Astronomy, University of California, Berkeley, CA 94720-3411, USA}
  \altaffiltext{9} {School of Mathematics and Physics, Queens University Belfast, Belfast BT7 1NN, UK}
\altaffiltext{10}  { Finnish Centre for Astronomy with ESO (FINCA), University of Turku, VŠisŠlŠntie 20, FI- 21500 Piikkiš,, Finland}
\altaffiltext{11}  { Tuorla Observatory, Department of Physics and Astronomy, University of Turku, VŠisŠlŠntie 20, FI-21500 Piikkiš, Finland}
\altaffiltext{12} {Steward Observatory, University of Arizona, 933 North Cherry Avenue, Tucson, AZ 85721, USA}
\altaffiltext{13}  { Department of Astronomy, University of Texas, Austin, TX 78712-0259, USA}
\altaffiltext{14}  { Columbia Astrophysics Laboratory, Columbia University, 550 West 120th Street, New York, NY 10027, USA}
\altaffiltext{15}  { 25 Nieuwland Science, University of Notre Dame, Notre Dame, IN 46556-5670, USA}
\altaffiltext{16}  { University of Bern, Center for Space and Habitability, Sidlerstrasse 5, CH-3012 Bern, Switzerland}
\altaffiltext{17}  { Department of Physics and Astronomy, Hofstra University, Hempstead, NY 11549, USA}
\altaffiltext{18}  { Space Telescope Science Institute, 3700 San Martin Drive, Baltimore, MD 21218, USA }
\altaffiltext{10}  { INAF-NA, Osservatorio Astronomico di Capodimonte, Salita Moiariello 16, 80131 Naples, Italy}
\altaffiltext{20}  { Supernova Ltd, OYV \#131, Northsound Road, Virgin Gorda VG 1150, Virgin Islands, UK}
\altaffiltext{21}  { Department of Physics, University of Hong Kong, Pokfulam Road, Hong Kong, PR China}
\altaffiltext{22}  { Observational Cosmology Laboratory Code 665, NASA Goddard Space Flight Center, Greenbelt, MD 20771, USA}
\altaffiltext{23}  { Department of Physics and Astronomy, Goucher College, 1021 Dulaney Valley Road, Baltimore, MD 21204, USA}
\begin{abstract}
We present imaging and spectroscopic observations with HST and VLT of the ring of SN 1987A from 1994 to 2014. After an almost exponential increase of the shocked emission from the hotspots up to day $\sim 8,000$ ($\sim 2009$), both this and the unshocked emission are now fading. From the radial positions of the hotspots we  see an acceleration of these up to $500-1000 \ \kms$, consistent with the highest spectroscopic shock velocities from the radiative shocks. In the most recent observations (2013 and 2014), we find several new hotspots {\it outside} the inner ring, excited by either X-rays from the shocks or by direct shock interaction. All of these observations indicate that the interaction with the supernova ejecta is now gradually dissolving the hotspots. We predict, based on the observed decay, that the inner ring will be destroyed by $\sim 2025$. 
 \end{abstract}

 \section{INTRODUCTION}
\label{sec-introd}
Supernova 1987A is unique: we can study in real time and with high spatial resolution physical processes that we can only infer for other
supernovae. Emission from SN 1987A included a flash of ultraviolet radiation in the first hours, thermal diffusion of radioactive energy input for weeks, and prompt reprocessing of radioactive power for decades \citep[e.g.,][]{McCray1993}. But now, conversion of the supernova's kinetic energy to heat through shock interactions dominates the radiative energy budget  \citep[][in the following L11]{Larsson2011}. 

The  ring system around SN 1987A was created by mass loss from the progenitor star in the 20,000 years before the explosion { \citep{Crotts1991}. The rings may have been created as a result of mass ejections from a rapidly rotating single star \citep{Chita2008}  or a merger of two massive stars \citep{Morris2009,Akashi2015}. Observations of the interaction of the ejecta  with the circumstellar medium can provide clues to the mass loss history and thereby the nature of the progenitor.

A blast wave, driven by the supernova ejecta,  initially with a velocity of  $\ga 35,000$ \kms \citep{Staveley-Smith1993}, 
slowed down to $\sim 4000$ \kms \ after interacting with the H II region created by the B3 Ia progenitor star in the swept-up red supergiant wind \citep{Chevalier1995,Dewey2012,Potter2014}. The first interaction with  the circumstellar ring occurred in 1995. This event was manifested by the appearance of the first ÒhotspotÓ \citep{Sonneborn1998,Lawrence2000}, where the blast wave entered a relatively dense clump of matter  and suddenly slowed to velocities of a few hundred \kms. 
Additional hotspots erupted in dense clumps of gas over the next few years, gradually involving the whole ring.
Emission at optical \citep{Groningsson2008b}, X-ray \citep{Maggi2012,Helder2013} and radio \citep{Ng2013} wavelengths have since then undergone a steep
rise.

We have followed the evolution of these shocks from the very beginning through regular monitoring with the Hubble Space Telescope (HST) and the Very Large Telescope (VLT).
In this Letter we report observations of the ring system over the last $\sim 20$ years, and show that the ring interaction has entered a new phase involving the final destruction of the ring but also interactions with material outside of this.

\section{OBSERVATIONS AND ANALYSIS}
\label{sec-obs}
 \begin{figure*}[ht!]
\centering
\includegraphics[width=11cm,angle=-90]{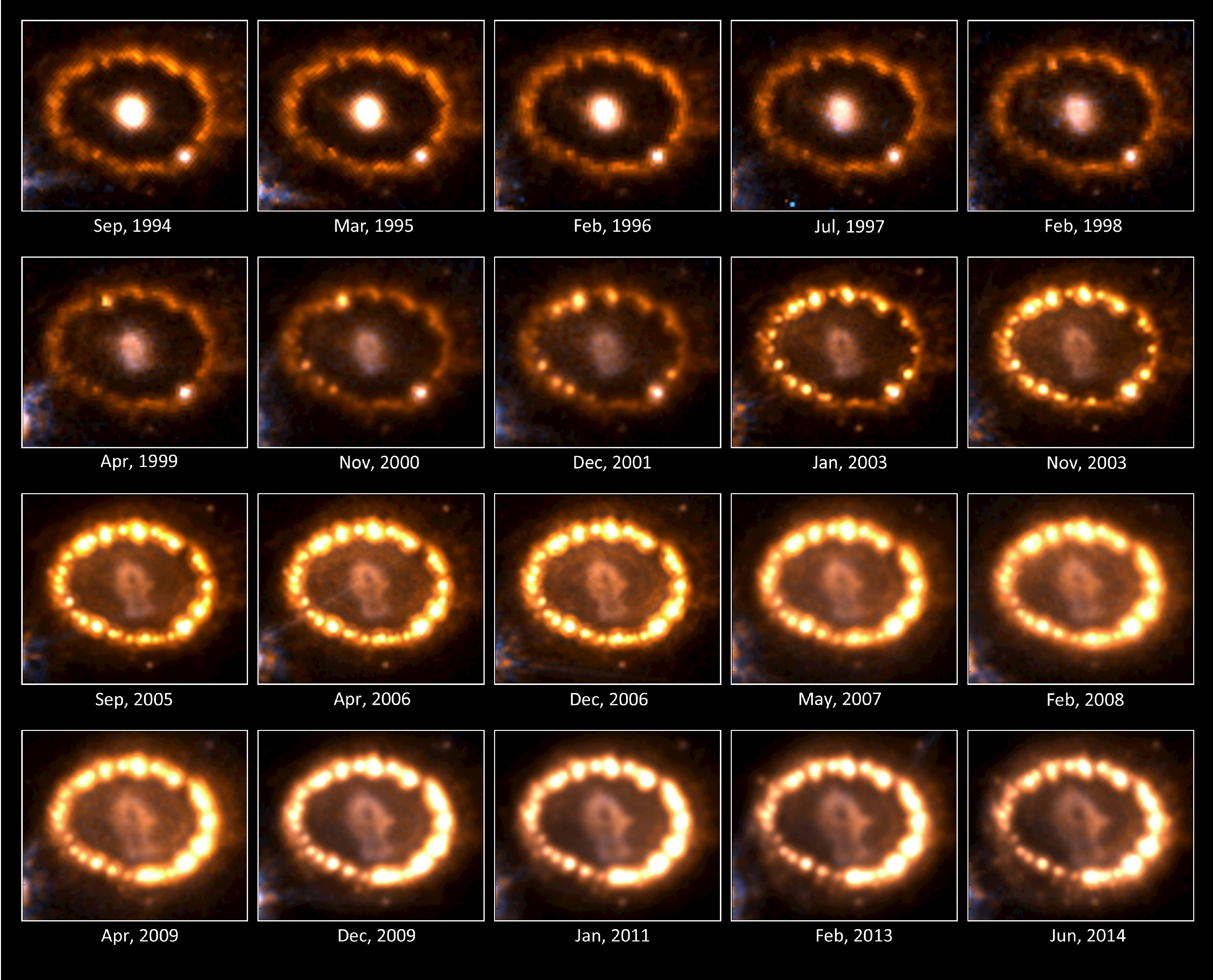}
\caption{ Evolution of the ring collision from 1994 to 2014 (days 2,270 - 9,975) from a combination of HST B- and R-band images. The brightness of the ring has been reduced by a factor of 20 by applying a mask to the images. This makes it possible to see the morphology of the ring at the same time as the faint ejecta and regions outside.  The new spots are better seen in the difference images in Fig. 4. The flux scale is otherwise the same for all panels.
The size of the field is $2.1\arcsec \times  1.8\arcsec$. North is up, east is left.}
\label{fig_hst_comp}
\end{figure*}

In order to make light curves of the ring we use all HST imaging observations in the R-band (F675W and F625W) and B-band (F439W, F435W and F438W) obtained since 1994. Details of all observations up to December 2009 are included in L11. Additional observations are given in Table \ref{obstab}. All images were processed in a standard way in order to remove cosmic rays and drizzled to combine dithered exposures (L11). Fig. \ref{fig_hst_comp} shows a time series of the combined 'R' and 'B' band images, illustrating the evolution of the supernova during the last 20 years. 
\begin{table}[h!]
\begin{center}
\caption{\label{obstab}\small{New UVES and HST observations. }}
\begin{tabular}{lcccc}
\hline\hline \\ [-8pt]
\multicolumn{1}{c}{Date} & \multicolumn{1}{c}{Epoch\footnote{Since 1987 Feb 23}} & 
& \multicolumn{1}{c}{Exposure}  & \\
 &  \multicolumn{1}{c}{(days)}  &  & \multicolumn{1}{c}{(s)}&    \\ \hline

UVES/VLT & & \multicolumn{1}{c}{Wavelength range }
&  & \multicolumn{1}{c}{Seeing}\\
 &   & \multicolumn{1}{c}{(nm)}   & & \multicolumn{1}{c}{(arcsec)}  \\ \hline
 2008-11-22 - & 7,943-  & 303-388+476-684 & 9000 & 0.8-1.0 \\ 
\hfill 2009-02-08 &\hfill 8,021   & 373-499+660-1060 & 11250 & 0.8-1.3 \\ 
\\
2010-10-18 -  & 8,638-  & 303-388+476-684 & 9000 & 0.7-1.1 \\ 
\hfill  2010-11-15&\hfill 8,666   & 373-499+660-1060 & 11250 & 0.6-1.1 \\ 
 \\
2011-10-05 -  & 8,990-  & 303-388+476-684 & 9000 & 0.8-1.1 \\ 
\hfill  2011-12-04&\hfill 9,050 & 373-499+660-1060 & 9000 & 0.6-1.4 \\ 
\\
2013-10-18 -  & 9,734-  & 303-388+476-684 & 8930 & 1.2-1.7 \\ 
\hfill 2013-12-20&\hfill 9,797  & 373-499+660-1060 & 8950 & 0.5-1.0 
 \\ \hline
 HST&& \multicolumn{1}{c}{Filter}&&\\ \hline
			
2011-01-05 &  8,717 & F438W &  1400&\\
                      &  & F625W &  1000&\\

2013-02-06 &  9,479 & F438W &  1200&\\
                      &  & F625W &  1200&\\

2014-06-15   & 9,973 & F438W &  1200& \\
                      &  & F625W &  1200& \\
2014-06-20  &  9,978 & F502N &  3080& \\	
2014-06-21 &  9,979 & F645N &  2880& \\
		   &  & F658N &  2880& \\

\hline
\end{tabular}
\end{center}
\end{table}

The light curve of the ring was measured using an aperture in the form of an elliptical annulus, with semi-major axis $0.5''$ ($1.1''$) for the inner (outer) boundary and an axial ratio of 0.78. Corrections were applied as described in L11 in order to account for differences between the different filters/instruments as well as the loss of charge transfer efficiency in the late WFPC2 images.  The resulting light curves are shown in the top panel of Fig. \ref{fig_lines}. 
\begin{figure}[th!]
\centering
\includegraphics[width=8.7cm]{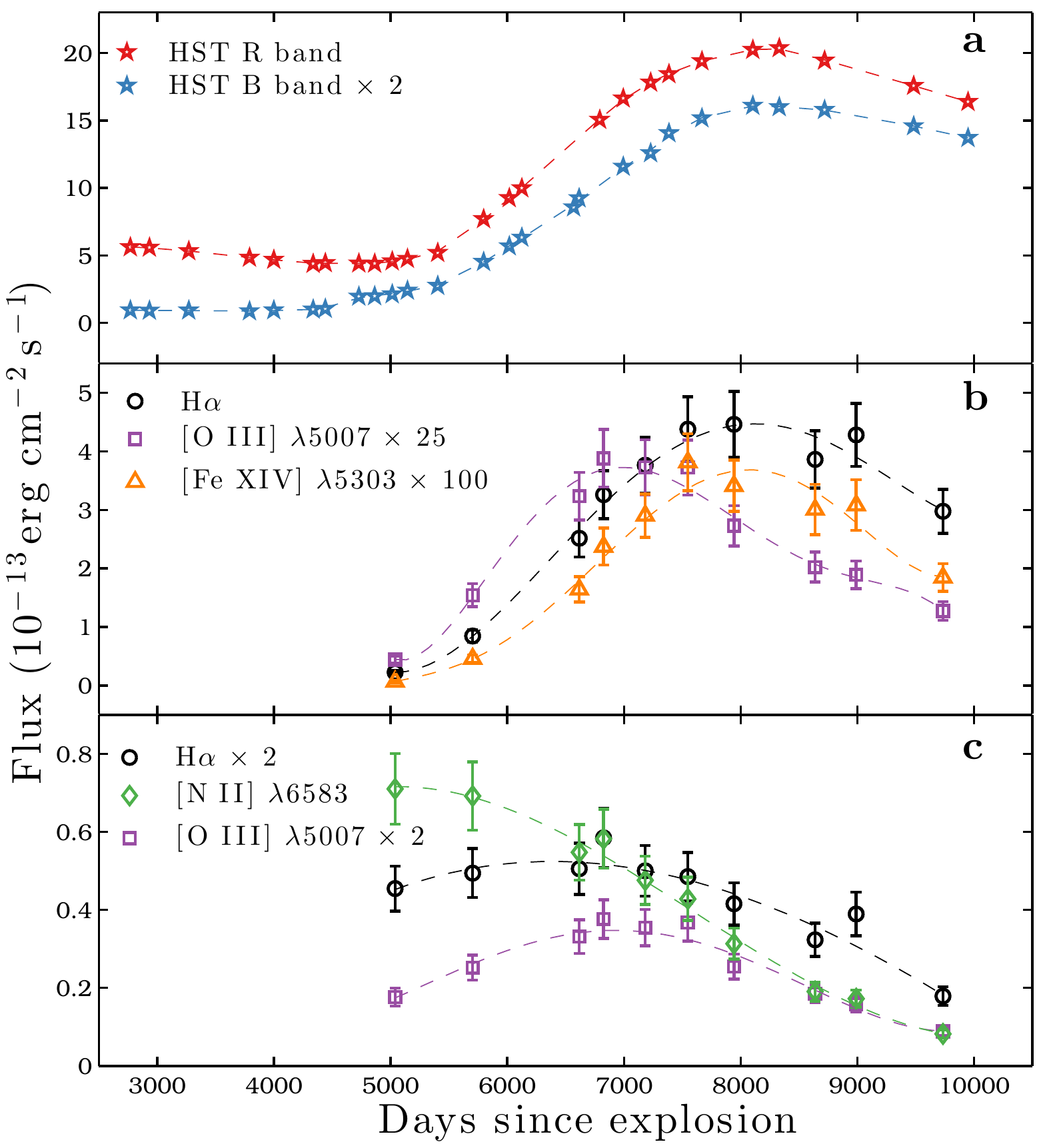}
\caption{Upper panel: R and B band light curves of the full ring from HST photometry. Middle panel: Light curve in individual emission lines from the shocked ring from spectroscopy with VLT/UVES, using a 0.8\arcsec slit. Note the different scales on the vertical axis in the panels and the scaling factors for the B-light curve and the lines. Bottom panel: The same for the narrow H$\alpha$, [N II] \wl 6583 and [O III] \wl 5007 lines from the unshocked clumps. }

\label{fig_lines}
\end{figure}

From the images we can also determine the positions and fluxes of the hotspots. To define the hotspots, we used an ACS image taken in the F625W filter  on December 6, 2006, which provides the best spatial resolution close to the time when the ring was brightest and when all hot spots are present (Fig. \ref{fig_hst_comp}). A two-dimensional Gaussian was fitted around each local maximum in the ring, resulting in the identification of  28 hotspots (see Fig. \ref{fig_spot_lc_r}). The distance of each hotspot from the center of the ring was measured by fitting a one-dimensional Gaussian to the intensity along a radial ray at the angular location of the spot in question.  In images taken prior to the onset of a hotspot, this method returns the radial location of the ring. Fig. \ref{fig_spot_lc_r} shows the time-evolution of the spot positions, given as a distance to the mean position of the ring before the appearance of the spots. 
\begin{figure}
\centering
\includegraphics[width=7.5cm]{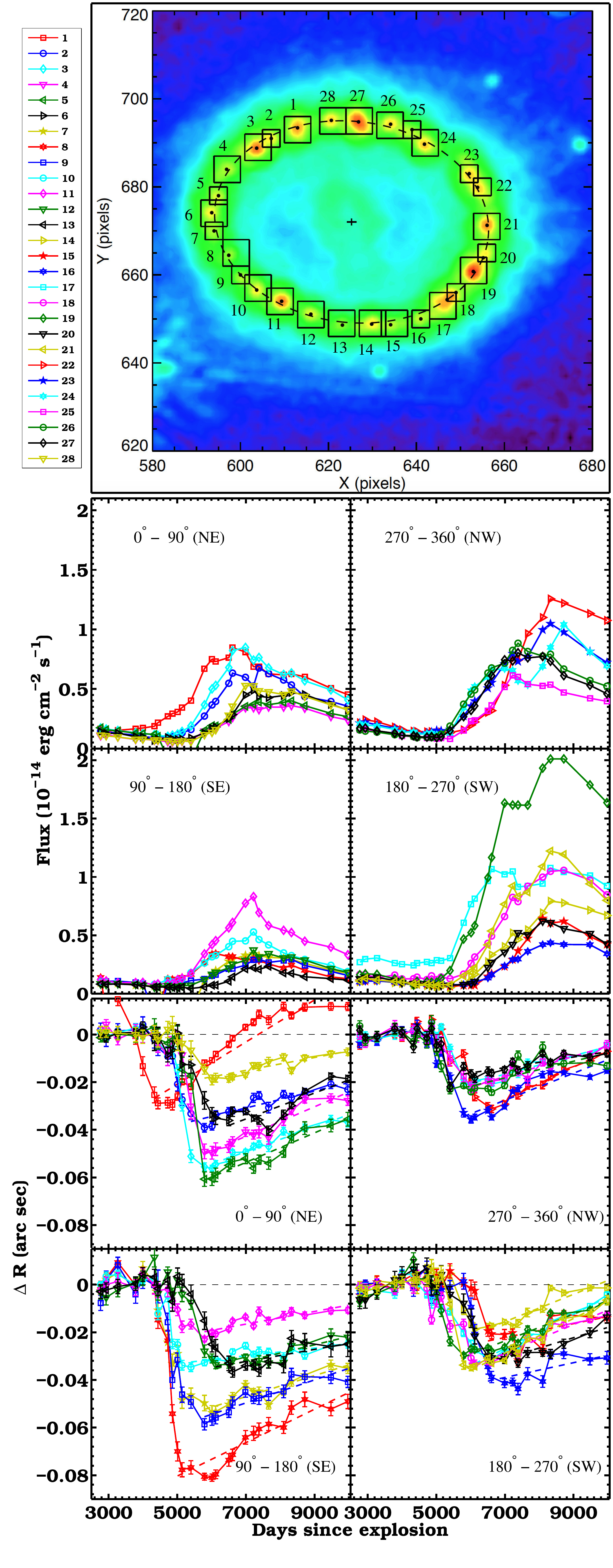}
\caption{Upper panel: Contour plot of the December 6, 2006 image.  
Boxed regions indicate where 2D Gaussian fits were applied, and the corresponding hotspot locations are identified with black dots.  
Middle panels: Evolution of the fluxes from the individual hotspots in the R band for the north-east (NE), north-west (NW), south-east (SE) and south-west (SW) quadrants of the ring numbered as in the image above.
Lower panels: Evolution of radii from the centre (in arc secs) of the different spots as a function of time for the different quadrants. The radial positions are calculated with respect to the average positions of the ring before the impact of the shock, seen as a drop at 5,000-6,000 days in the radial positions. The dashed lines give a linear least square fit to the data after the impacts of the different spots used for the velocity determinations.}
\label{fig_spot_lc_r}
\end{figure}

The velocities for each spot were determined from a linear fit to the positions after the impact, seen as a sudden decrease in the radius. For some of the spots there are indications of a decreasing
velocity after $\sim 8,000$ days. Given the errors in the measurements we have, however, limited the analysis to a linear least square fit from the time of the impact to the last observation. The resulting velocities range between 180 $\kms$ and 950 $\kms$ with a mean at $\sim$540 $\kms$. We estimate the errors in the velocities to be $\sim 50 \ \kms$. 

The flux from each spot was determined from the FWHM and peak of the Gaussian. The result for each epoch is shown in the middle panel in Fig  \ref{fig_spot_lc_r}, corrected for differences in filters and instruments in the same way as the total light curve.

To complement the imaging high resolution spectra were obtained at regular intervals with VLT/UVES between December 2000 and December 2013. These spectra allow us to separate the contribution from shocked and unshocked gas. Details of observations are given in Gr{\"o}ningsson et al. (2008b) for observations up to 2007 and later in Table \ref{obstab}. Reductions are similar, using the updated UVES pipeline.
The $0.8"$ slit was centred at the supernova and oriented at a P.A. of 30${}^ \circ$. 

We estimate the accuracy of the fluxes from the UVES spectra by comparing them to those from HST imaging in R and B bands, taking the slit width and seeing into account.  
We find that the UVES fluxes are within 20\% of the HST fluxes, and that they also follow the same trend in time (Fig. \ref{fig_lines}). The errors in the fluxes for most of the lines are dominated by this systematic error. However, for H$\alpha$ at the last epochs the separation between the increasingly faint narrow component from the much brighter shocked component becomes increasingly difficult. This may introduce a systematic underestimate of the flux from the unshocked spots by up to $\sim 50\%$  for the last observations, resulting in a similar increase in these fluxes. The H$\alpha$ light curve would then be marginally flatter, but the general strong decrease should not be affected by this.

\section{RESULTS}
\label{sec-res}

The light curves of the ring 
in the upper panel of Fig. \ref{fig_lines} are characterized by a slowly decreasing part up to day $\sim 5,000$, followed by a fast rise, which reached its peak at day $\sim 8,000$, after which they have decayed. 
The B-band is dominated by emission lines from
H$\gamma$, H$\delta$ and [S II] \wl 4069, 
while the R-band is dominated by H$\alpha$, [N II] \wll 6548, 6583, and [O I] \wll 6300, 6364. 
The initial decay is a result of the recombination and cooling after the initial ionization from the shock breakout \citep{Lundqvist1996}.
The strong increase in the flux began as the outward shock interacted with the whole ring, producing the hotspots, which brightened with time. This phase ended $\sim 8,000$ days after the explosion and the hotspots are now fading in our observations that extend to 9,975 days. 

 The turn-on of the hot spots occurred between 5000-6000 days at the same time as the emission peaks moved inward (Fig. \ref{fig_spot_lc_r}). These dense 'protrusions' may have recombined at early time due to their high density and only became visible when hit by the shock. The flux from the hotspots peaked earliest on the east side of the ring.

The high-resolution spectroscopy from the UVES observations allows us to separate the contribution to the light curve from the unshocked ring (emitting narrow lines with FWHM $\sim 10$ \kms \citep{Groningsson2008a})
and the shocked hotspots (emitting lines with typical velocities $\sim 300 \ \kms$).
In the middle panel of Fig.  \ref{fig_lines} we show the light curves of three different lines from the shocked gas (H$\alpha$, [Fe XIV] \wl 5303 and [O III] \wl 5007), which all show a similar evolution as the HST light curves. 
In the lower panel we show the fluxes of the strongest narrow lines from the unshocked gas (H$\alpha$, [N II] \wl 6583 and [O III] \wl 5007). Up to $\sim$7,000 days the unshocked [O III] and H$\alpha$ lines increase in flux, caused by pre-ionization by the soft X-rays. After this epoch all three lines, however, decline more steeply than the lines from the shocked gas.

From the UVES observations of H$\alpha$ we also find that the shocks are radiative up to at least $\sim 700 \ \kms$, and possibly up to $\sim 1,000 \ \kms$, depending on the angle of the shocks relative to the line of sight \citep[][Migotto et al., 2015, in prep.]{Groningsson2008b}. Using the cooling time from \cite{Groningsson2008b} this indicates a density of up to $\sim 6\times 10^4 \ccm$ in the clumps. This can be compared to the velocities derived from the proper motions in Sect. \ref{sec-obs}, where the average was found to be $\sim 540 \ \kms$.  This directly reflects the propagation of the shocks through the clumps and and concomitant acceleration of the post-shock gas. Although the radial and proper motion velocities are in perpendicular directions, the $43 \degr$ inclination
of the ring results in a similar correction, assuming that the expansion is radial from the explosion centre. These two methods therefore result in similar shock velocities. 

The most interesting recent changes are shown in Fig.  \ref{fig_diff}, with the emergence of new, faint spots, as well as diffuse emission, outside the ring. 
 \begin{figure*}[ht!]
\centering
\includegraphics[width=4cm]{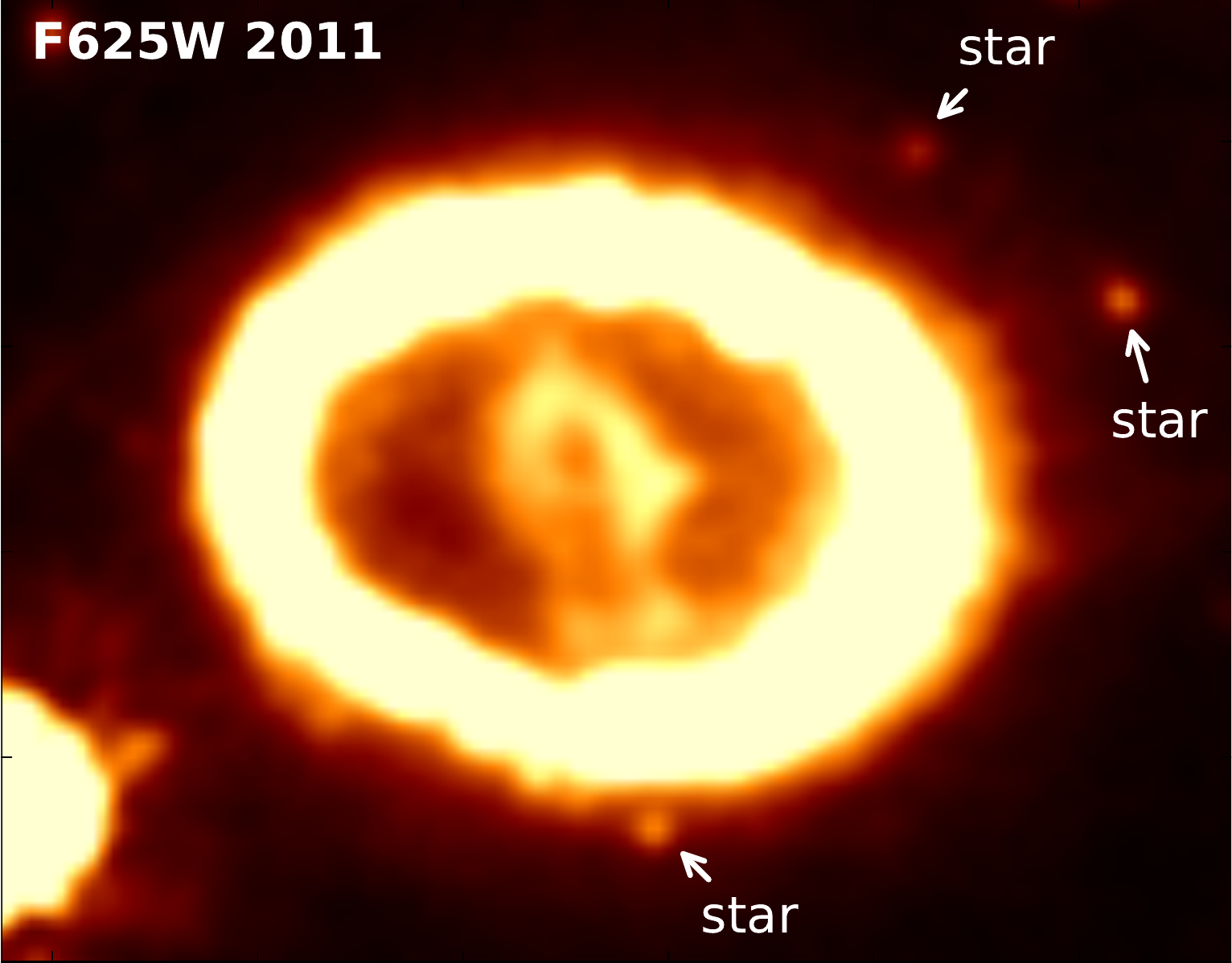}
\includegraphics[width=4cm]{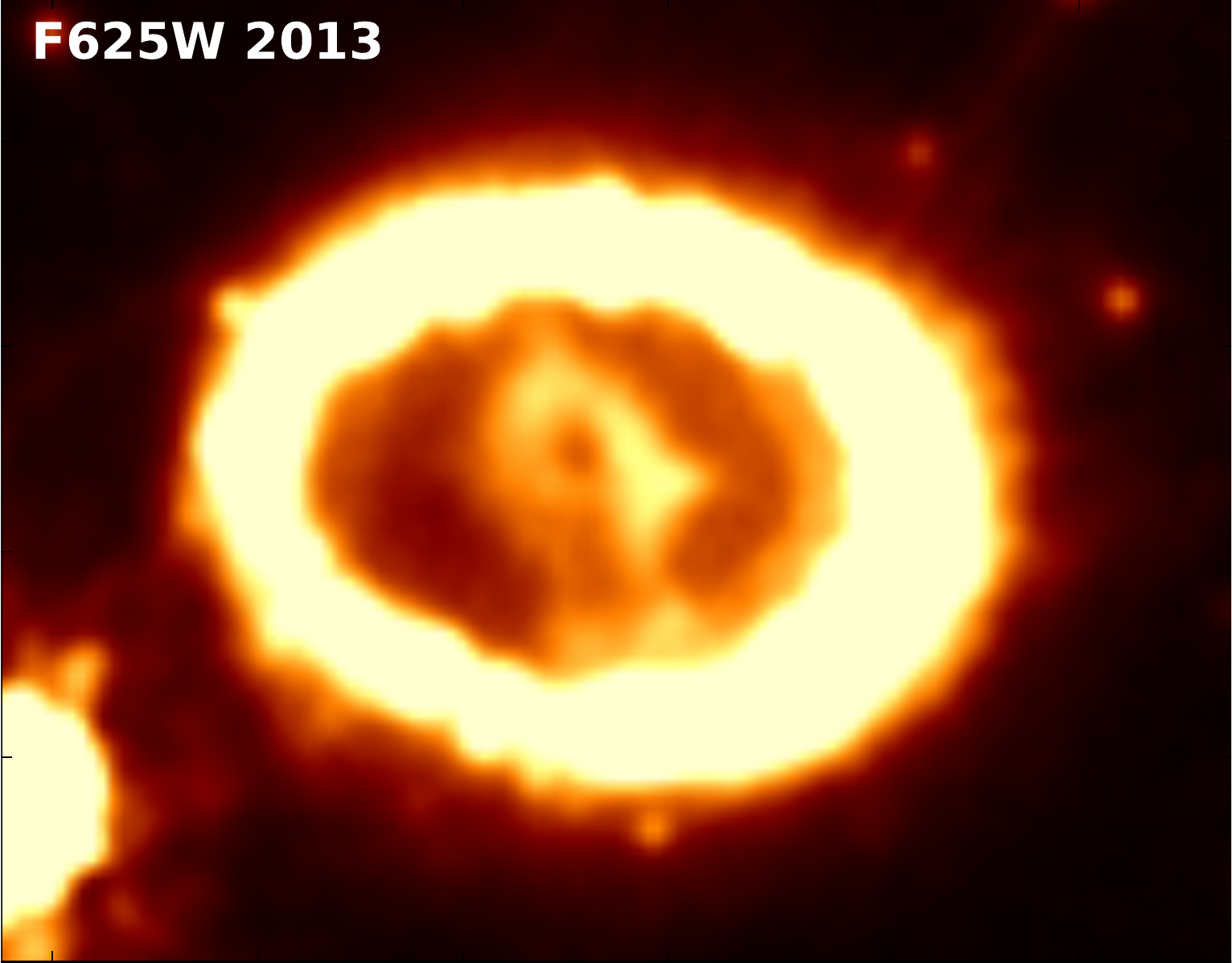}
\includegraphics[width=4.53cm]{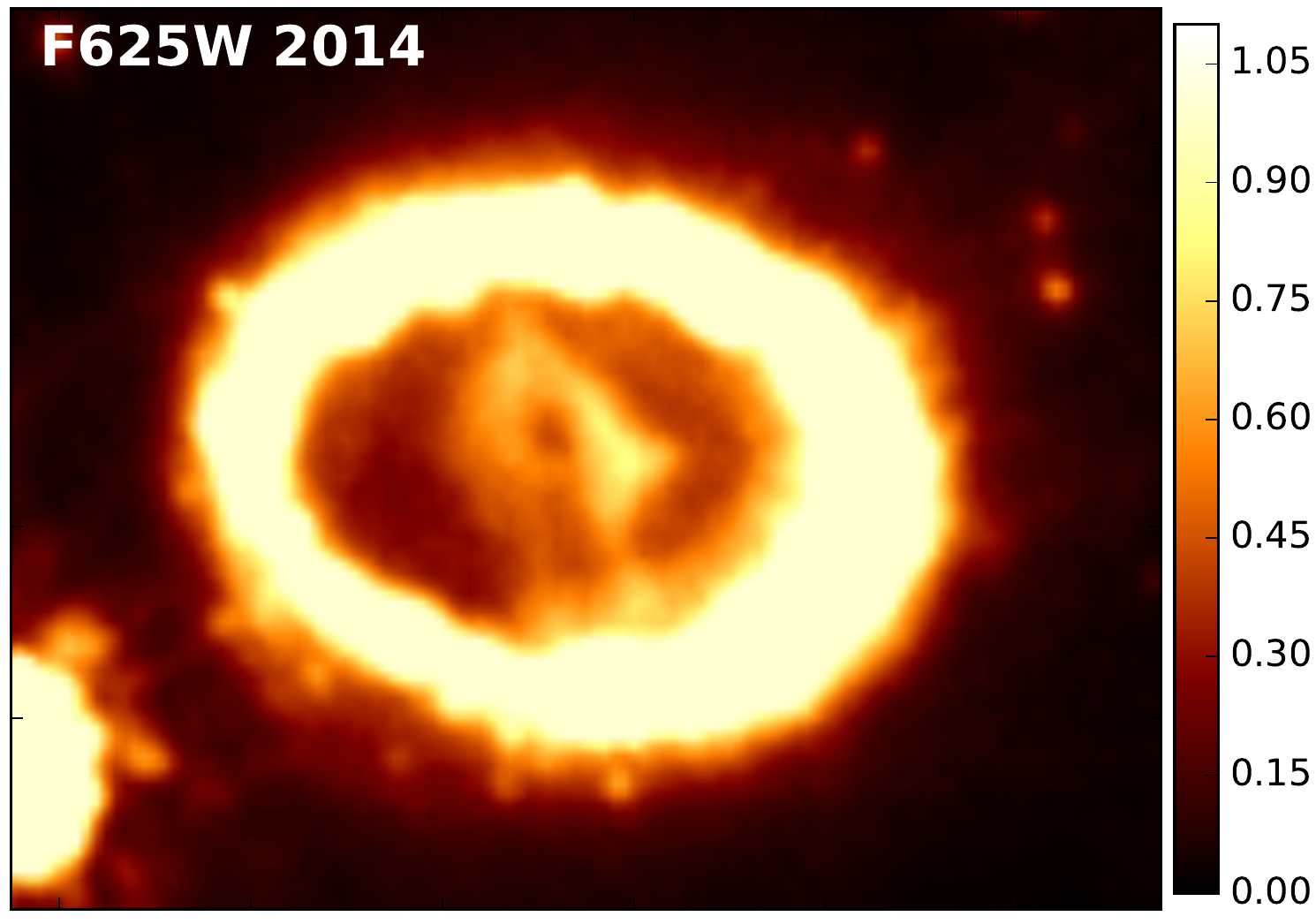}\\
\includegraphics[width=4cm]{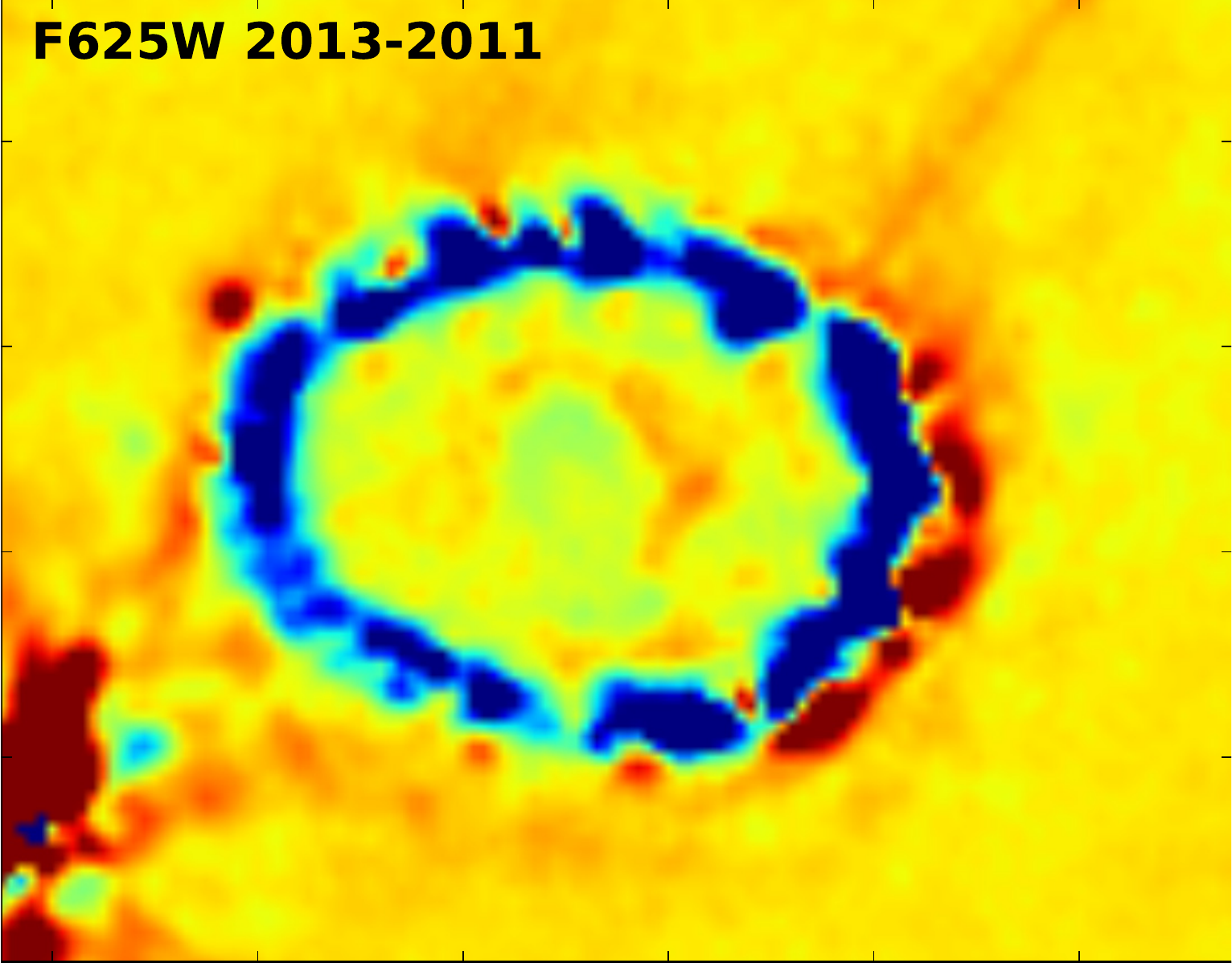}
\includegraphics[width=4cm]{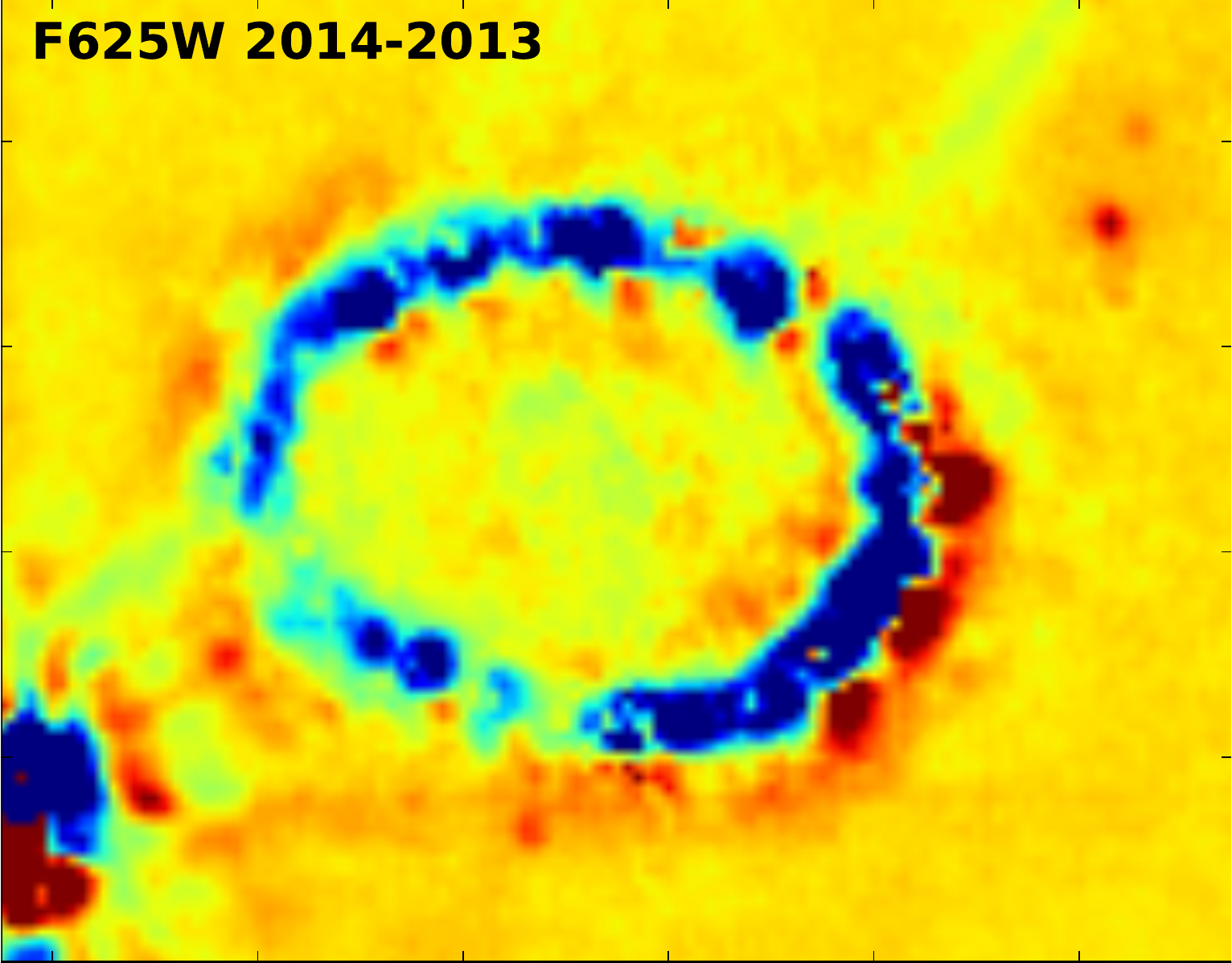}
\includegraphics[width=4.53cm]{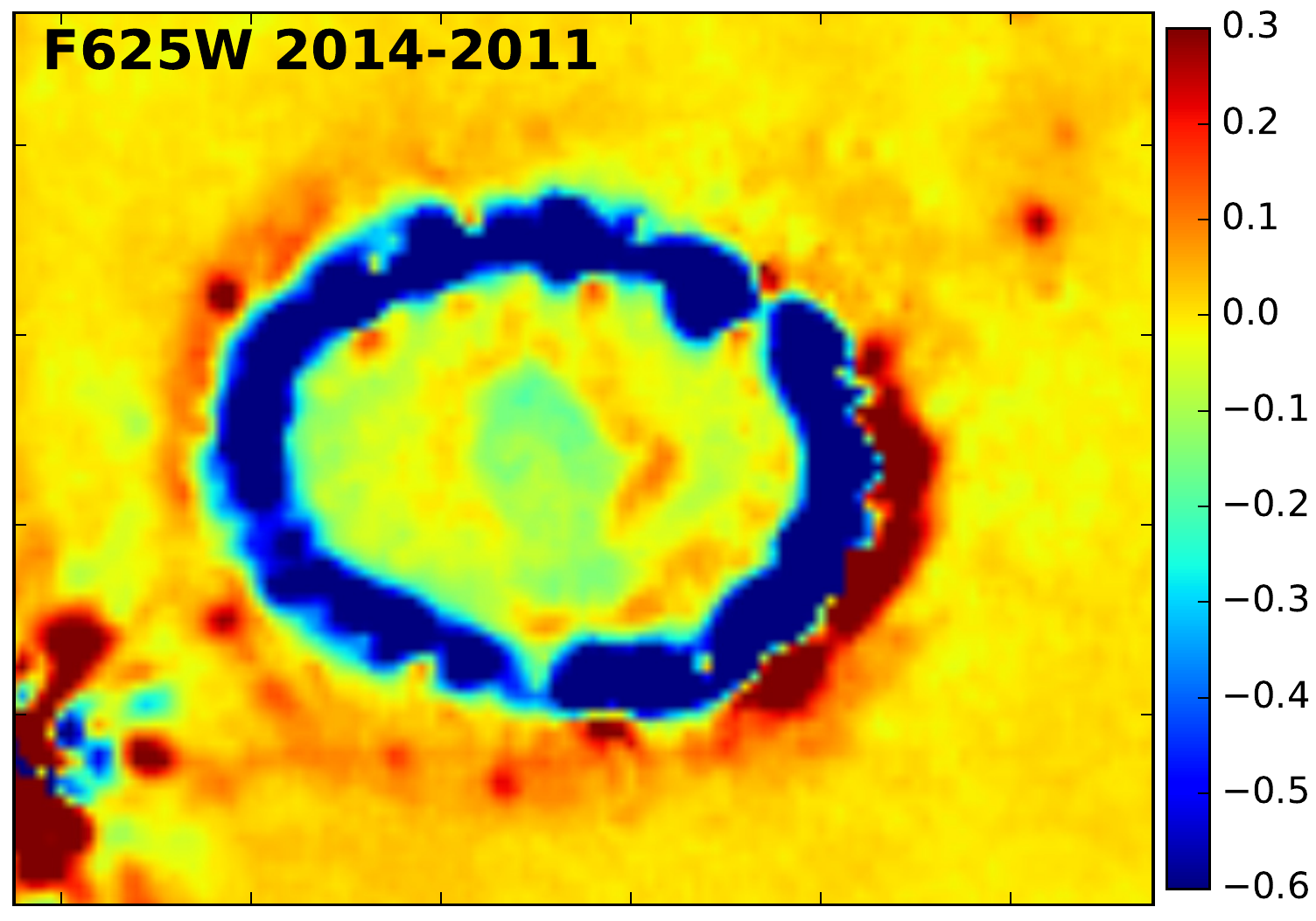}\\
\includegraphics[width=4cm]{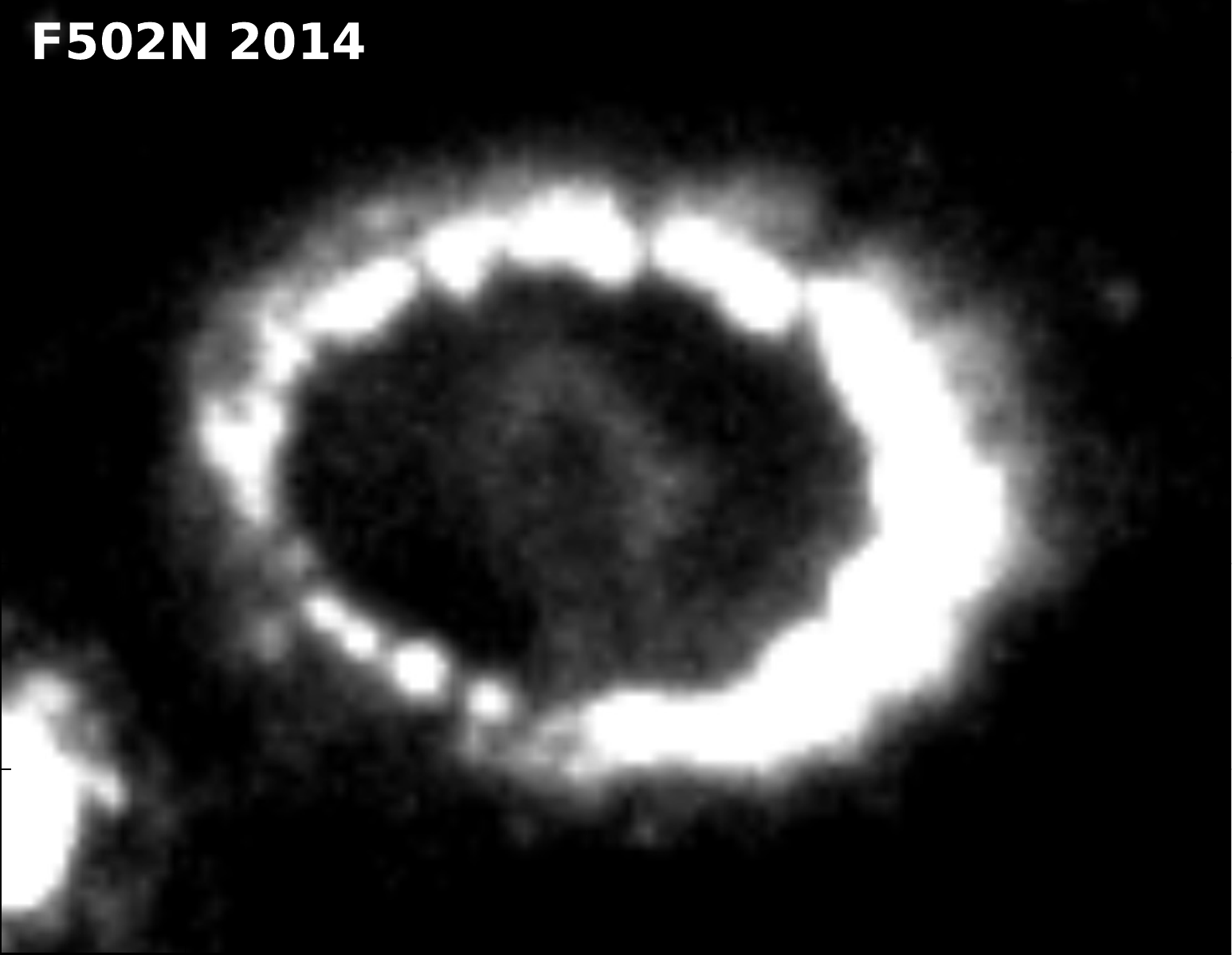}
\includegraphics[width=4cm]{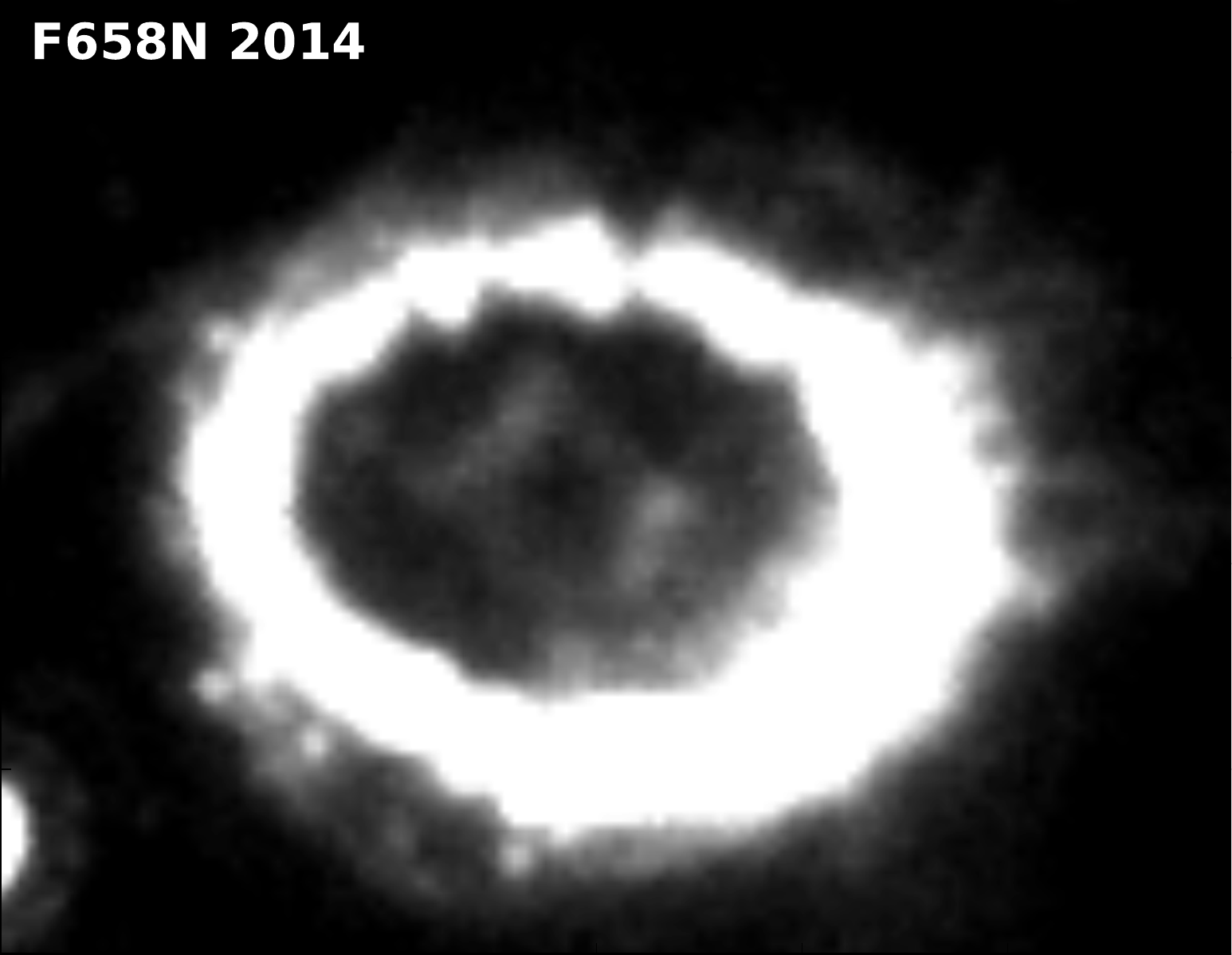}
\includegraphics[width=4cm]{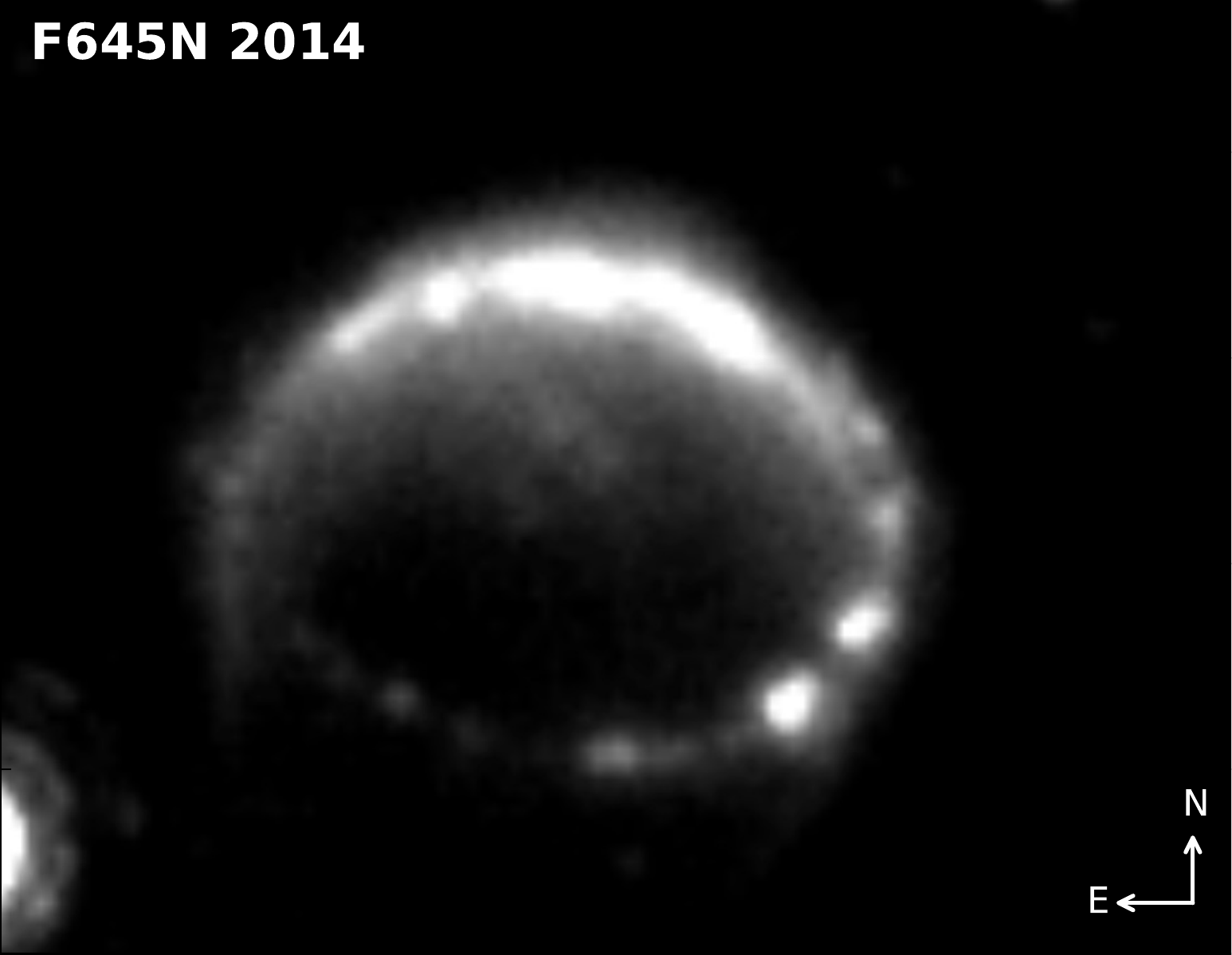}
\hspace{0.3cm}
\caption{Upper panel: HST images of SN 1987A  with WFC3 in the F625W filter on days 8,717 (2011-01-05), 9,480 (2013-02-06),  and 9,975 (2014-06-15). Middle panels: Difference images of 2013-2011, 2014-2013 and 2014-2011, respectively. Blue means fainter and red brighter. Inside the ring the asymmetric ejecta can be seen. Note the gradual appearance of several small spots, as well as diffuse emission, outside the inner ring in the south-east. New spots can also be seen in the north-east, as well as to the north-west. Note that the two new spots in the north-west do not coincide with the stars labelled in the upper left image. The residuals in the south-east corner of the difference images are due to a saturated star and the radial streak north-west of the ring in the two first difference panels is due to a diffraction spike of a star outside the field shown.  Lower panels: Narrow band images in the F502N ([O III]), F658N ([N II]) and F645N (continuum) filters. The north-south asymmetries in these images are due to lines being shifted in/out of the narrow band filters.The field size is $3.0\arcsec \times  2.4\arcsec$. }
\label{fig_diff}
\end{figure*}
These new spots are most clearly seen in the final image from 2014, but with hindsight, some of these are faintly visible in the 2013 image. The new hotspots are a factor of 10--20 fainter than the old hotspots. The new spots, as well as diffuse emission outside the ring, can also be seen in the narrow-band images in Fig. \ref{fig_diff}. It is interesting that most of the new spots are in the same region as where the faintest  hotspots in the ring are present. The diffuse emission is likely to be a combination of H$\alpha$ from the reverse shock region \citep{France2015} and gas producing line emission outside the ring. The latter may be a result of photoionization of the wind from the red supergiant stage of the progenitor by X-rays from the shock.

\section{DISCUSSION}
\label{sec-disc}

To interpret these observations we consider the ram pressure acting on the clumps, given by $P \propto \rho_{\rm ejecta} V^2  \propto V^{-n+2} t^{-3} \propto t^{n-5}$, where $n$ is the power law index of the ejecta density
profile, $ \rho_{\rm ejecta} \propto V^{-n}$, $V$ is the ejecta velocity at the reverse shock, assumed to be at a constant radius, and $t$ is the time since explosion. The pressure will increase with time as long as the reverse shock is in the steep part of the ejecta density profile. This pressure drives transmitted shocks into clumps of gas in the circumstellar ring. 

If the density of a clump is sufficiently high, radiative cooling will cause the clump to collapse to an even higher density. The result will be an optically emitting hotspot \citep{Pun2002}. If instead, the density of a
clump falls below the threshold density for radiative cooling, the gas behind the transmitted shock will radiate X-rays and very little optical radiation. 
This threshold density is a sensitive function of the velocity, $V_{\rm clump}$, of the shock entering the clump.
Thus, at any given time, the requirement for radiative cooling acts as a high contrast filter, limiting optical radiation to only the denser clumps. This may explain the still increasing soft X-ray emission \citep{Maggi2012,Helder2013}, in contrast to the optical. { \cite{Dewey2012} find from modeling of the X-ray observations a filling factor of $\sim 30 \%$ for the dense clumps.

If the radiative shocks can be characterized by one velocity the luminosity will depend on time as $L \approx \rho_{\rm clump} V_{\rm clump}^3 A \propto  \rho_{\rm clump}^{-1/2}  t^{(3n-15)/2} A \propto  t^{5.4} A$, where $\rho_{\rm clump}$ is the density and $A$ is the cross section of the clumps and we have assumed $n \approx 8.6$ \citep{Shigeyama1990}. Explosion models show that the reverse shock is still likely to be in the steep density part of the ejecta  \citep{Dewey2012,Fransson2013}. Further, the velocity for which the shocks are radiative is not likely to decrease because the density is increasing as long as $n > 3$. The most likely reason for a decrease in the luminosity is therefore that the area of the radiative shocks is decreasing. Alternatively, or in addition, as the driving pressure increases, the transmitted shock velocity may exceed the threshold for radiative cooling. Likewise, the threshold density for a clump to accommodate a radiative shock will increase with time. In either case it will lead to the clumps  getting dissolved.

From the fading and acceleration of the hotspots we conclude that the clumps are destroyed by the blast wave, in agreement with hydrodynamic simulations \citep{Borkowski1997,Pun2002}. The fast decay of the narrow lines from the unshocked gas can be explained as a result of the non-radiative shocks, which traverse lower density gas, replacing the narrow line emission with soft X-ray emission, in combination with a decreasing emission measure of the pre-shock gas in the clumps.

The range of densities in the unshocked ring has been estimated to $1 \times 10^2 \ \ccm$ up to $3 \times 10^4 \ \ccm$ \citep{Mattila2010}, while the shock speed from the X-ray imaging is $\sim 1,820 \ \kms$ \citep{Maggi2012,Helder2013}, coming from shocks propagating in the low-density component of the gas. Assuming that the interaction with the ring started on day 5,600 \citep{Helder2013}, the blast wave has expanded by $\sim 7 \times 10^{16}$ cm up to day 9,975, or $\sim 10 \%$ of the radius of the ring. This is of the same order as the distance of the new hotspots outside the ring. It is therefore conceivable that the new hotspots are a result of either the interaction of the blast wave with very dense clumps embedded in gas with density $\sim 10^3  \ \ccm$  outside the ring, and close to the equatorial plane, or pre-ionization by the X-rays. The fact that the main part of the ring has faded most in the south-east, where the majority of the new spots are seen supports the former interpretation.

The time scale for the complete destruction of the ring depends on the mass, geometry and density distribution in the clumps \citep{Pun2002}. The mass estimate for the ring, $5.8 \times 10^{-2} \Mo$ \citep{Mattila2010}, only refers to the mass ionized by the initial shock breakout, and the mass of the hotspots and gas outside of the ring could be considerably larger. However, extrapolating the light curves in Fig. \ref{fig_lines} shows that the ring should fade away between $\sim 2020$ (based on the
individual lines) and $\sim$ 2030 (based on the HST light curves). The former estimate is likely to be more reliable, as the lines isolate the emission from the shocked ring component. The hotspots will be gradually destroyed by instabilities and conduction by the hot surrounding gas \citep{Borkowski1997,Pun2002}. The shocked gas which has cooled and increased its density by a factor $100-1000$ may fragment and survive as overdense 'nodules' in the hot gas, unless heat conduction destroys them \citep{Poludnenko2004,Raga2007,Silvia2010}. This depends sensitively on the magnetic field and its orientation.  Heat conduction or X-ray irradiation may then provide some energy input to the surviving fragments, but at a low level. The cloud crushing time is $\sim 2 r_{\rm clump}/V_{\rm clump}$ \citep[e.g.,][]{Poludnenko2004}. With a decay time scale of $\sim 2000$ days and shock velocity of $\sim 500$ \kms  \ this corresponds to a clump  diameter of $\sim 0.9\times 10^{16}$ cm. For the individual spots in Sect. \ref{sec-obs} we find a FWHM of 1.5 pixels, consistent with being unresolved with the ACS. This corresponds to a diameter $< 0.04\arcsec$, or $< 3\times 10^{16}$ cm.  We do not find any evidence of a change in the lateral direction within $\pm 2 \degr$.

The structure outside the inner ring is complex with a possible hour-glass shape \citep{Sugerman2005,France2015}, which will now be possible to probe. As the shock progresses beyond the circumstellar ring, it will trace the history of mass loss from the supernova's progenitor, revealing the distribution of gas that is now unseen, and providing useful information to discriminate among different models for the progenitor of SN 1987A.

The lower general density will result in a mainly adiabatic shock wave, except for clumps with densities comparable to the inner ring. However, the new hotspots we found probably only represent a small fraction of the mass outside the ring. We expect SN 1987A will become more thoroughly X-ray dominated as the youngest supernova remnant evolves.

\acknowledgments
 We are grateful to the referee for detailed comments. This work was supported by the Swedish Research Council and the Swedish National Space Board, NASA grant NNX12AF90G, NSF grant AST-1109801. Support for the HST observing program was provided by NASA through a grant from the Space Telescope Science Institute, which is operated by the Association of Universities for Research in Astronomy, Inc. Partially based on observations collected at the European Southern
Observatory, Chile (ESO Programmes 080.D-0727, 082.D-0273, 086.D-0713,
088.D-0638, 090.D-645, 092.D-0119, 094.D-0505).


\begin{thebibliography}{28}
\expandafter\ifx\csname natexlab\endcsname\relax\def\natexlab#1{#1}\fi

\bibitem[{{Akashi} {et~al.}(2015){Akashi}, {Sabach}, {Yogev}, \&
  {Soker}}]{Akashi2015}
{Akashi}, M., {Sabach}, E., {Yogev}, O., \& {Soker}, N. 2015, ArXiv e-prints 1502.05541

\bibitem[{{Borkowski} {et~al.}(1997){Borkowski}, {Blondin}, \&
  {McCray}}]{Borkowski1997}
{Borkowski}, K.~J., {Blondin}, J.~M., \& {McCray}, R. 1997, \apj, 477, 281

\bibitem[{{Chevalier} \& {Dwarkadas}(1995)}]{Chevalier1995}
{Chevalier}, R.~A., \& {Dwarkadas}, V.~V. 1995, \apjl, 452, L45

\bibitem[{{Chita} {et~al.}(2008){Chita}, {Langer}, {van Marle},
  {Garc{\'{\i}}a-Segura}, \& {Heger}}]{Chita2008}
{Chita}, S.~M., {Langer}, N., {van Marle}, A.~J., {Garc{\'{\i}}a-Segura}, G.,
  \& {Heger}, A. 2008, \aap, 488, L37

\bibitem[{{Crotts} \& {Heathcote}(1991)}]{Crotts1991}
{Crotts}, A.~P., \& {Heathcote}, S.~R. 1991, \nat, 350, 683

\bibitem[{{Dewey} {et~al.}(2012){Dewey}, {Dwarkadas}, {Haberl}, {Sturm}, \&
  {Canizares}}]{Dewey2012}
{Dewey}, D., {Dwarkadas}, V.~V., {Haberl}, F., {Sturm}, R., \& {Canizares},
  C.~R. 2012, \apj, 752, 103

\bibitem[{{France} {et~al.}(2015){France}, {McCray}, {Fransson}, {Larsson},
  {Frank}, {Burrows}, {Challis}, {Kirshner}, {Chevalier}, {Garnavich}, {Heng},
  {Lawrence}, {Lundqvist}, {Smith}, \& {Sonneborn}}]{France2015}
{France}, K., {et~al.} 2015, \apjl, 801, L16

\bibitem[{{Fransson} {et~al.}(2013){Fransson}, {Larsson}, {Spyromilio},
  {Chevalier}, {Gr{\"o}ningsson}, {Jerkstrand}, {Leibundgut}, {McCray},
  {Challis}, {Kirshner}, {Kjaer}, {Lundqvist}, \& {Sollerman}}]{Fransson2013}
{Fransson}, C., {et~al.} 2013, \apj, 768, 88

\bibitem[{{Gr{\"o}ningsson} {et~al.}(2008{\natexlab{a}}){Gr{\"o}ningsson},
  {Fransson}, {Leibundgut}, {Lundqvist}, {Challis}, {Chevalier}, \&
  {Spyromilio}}]{Groningsson2008b}
{Gr{\"o}ningsson}, P., {Fransson}, C., {Leibundgut}, B., {Lundqvist}, P.,
  {Challis}, P., {Chevalier}, R.~A., \& {Spyromilio}, J. 2008{\natexlab{a}},
  \aap, 492, 481

\bibitem[{{Gr{\"o}ningsson} {et~al.}(2008{\natexlab{b}}){Gr{\"o}ningsson},
  {Fransson}, {Lundqvist}, {Lundqvist}, {Leibundgut}, {Spyromilio},
  {Chevalier}, {Gilmozzi}, {Kj{\ae}r}, {Mattila}, \&
  {Sollerman}}]{Groningsson2008a}
{Gr{\"o}ningsson}, P., {et~al.} 2008{\natexlab{b}}, \aap, 479, 761

\bibitem[{{Helder} {et~al.}(2013){Helder}, {Broos}, {Dewey}, {Dwek}, {McCray},
  {Park}, {Racusin}, {Zhekov}, \& {Burrows}}]{Helder2013}
{Helder}, E.~A., {et~al.} 2013, \apj, 764, 11

\bibitem[{{Larsson} {et~al.}(2011){Larsson}, {Fransson}, {{\"O}stlin},
  {Gr{\"o}ningsson}, {Jerkstrand}, {Kozma}, {Sollerman}, {Challis}, {Kirshner},
  {Chevalier}, {Heng}, {McCray}, {Suntzeff}, {Bouchet}, {Crotts}, {Danziger},
  {Dwek}, {France}, {Garnavich}, {Lawrence}, {Leibundgut}, {Lundqvist},
  {Panagia}, {Pun}, {Smith}, {Sonneborn}, {Wang}, \& {Wheeler}}]{Larsson2011}
{Larsson}, J., {et~al.} 2011, \nat, 474, 484

\bibitem[{{Lawrence} {et~al.}(2000){Lawrence}, {Sugerman}, {Bouchet}, {Crotts},
  {Uglesich}, \& {Heathcote}}]{Lawrence2000}
{Lawrence}, S.~S., {Sugerman}, B.~E., {Bouchet}, P., {Crotts}, A.~P.~S.,
  {Uglesich}, R., \& {Heathcote}, S. 2000, \apjl, 537, L123

\bibitem[{{Lundqvist} \& {Fransson}(1996)}]{Lundqvist1996}
{Lundqvist}, P., \& {Fransson}, C. 1996, \apj, 464, 924

\bibitem[{{Maggi} {et~al.}(2012){Maggi}, {Haberl}, {Sturm}, \&
  {Dewey}}]{Maggi2012}
{Maggi}, P., {Haberl}, F., {Sturm}, R., \& {Dewey}, D. 2012, \aap, 548, L3

\bibitem[{{Mattila} {et~al.}(2010){Mattila}, {Lundqvist}, {Gr{\"o}ningsson},
  {Meikle}, {Stathakis}, {Fransson}, \& {Cannon}}]{Mattila2010}
{Mattila}, S., {Lundqvist}, P., {Gr{\"o}ningsson}, P., {Meikle}, P.,
  {Stathakis}, R., {Fransson}, C., \& {Cannon}, R. 2010, \apj, 717, 1140

\bibitem[{{McCray}(1993)}]{McCray1993}
{McCray}, R. 1993, \araa, 31, 175

\bibitem[{{Morris} \& {Podsiadlowski}(2009)}]{Morris2009}
{Morris}, T., \& {Podsiadlowski}, P. 2009, \mnras, 399, 515

\bibitem[{{Ng} {et~al.}(2013){Ng}, {Zanardo}, {Potter}, {Staveley-Smith},
  {Gaensler}, {Manchester}, \& {Tzioumis}}]{Ng2013}
{Ng}, C.-Y., {Zanardo}, G., {Potter}, T.~M., {Staveley-Smith}, L., {Gaensler},
  B.~M., {Manchester}, R.~N., \& {Tzioumis}, A.~K. 2013, \apj, 777, 131

\bibitem[{{Poludnenko} {et~al.}(2004){Poludnenko}, {Frank}, \&
  {Mitran}}]{Poludnenko2004}
{Poludnenko}, A.~Y., {Frank}, A., \& {Mitran}, S. 2004, \apj, 613, 387

\bibitem[{{Potter} {et~al.}(2014){Potter}, {Staveley-Smith}, {Reville}, {Ng},
  {Bicknell}, {Sutherland}, \& {Wagner}}]{Potter2014}
{Potter}, T.~M., {Staveley-Smith}, L., {Reville}, B., {Ng}, C.-Y., {Bicknell},
  G.~V., {Sutherland}, R.~S., \& {Wagner}, A.~Y. 2014, \apj, 794, 174

\bibitem[{{Pun} {et~al.}(2002){Pun}, {Michael}, {Zhekov}, {McCray},
  {Garnavich}, {Challis}, {Kirshner}, {Baron}, {Branch}, {Chevalier},
  {Filippenko}, {Fransson}, {Leibundgut}, {Lundqvist}, {Panagia}, {Phillips},
  {Schmidt}, {Sonneborn}, {Suntzeff}, {Wang}, \& {Wheeler}}]{Pun2002}
{Pun}, C.~S.~J., {et~al.} 2002, \apj, 572, 906

\bibitem[{{Raga} {et~al.}(2007){Raga}, {Esquivel}, {Riera}, \&
  {Vel{\'a}zquez}}]{Raga2007}
{Raga}, A.~C., {Esquivel}, A., {Riera}, A., \& {Vel{\'a}zquez}, P.~F. 2007,
  \apj, 668, 310

\bibitem[{{Shigeyama} \& {Nomoto}(1990)}]{Shigeyama1990}
{Shigeyama}, T., \& {Nomoto}, K. 1990, \apj, 360, 242

\bibitem[{{Silvia} {et~al.}(2010){Silvia}, {Smith}, \& {Shull}}]{Silvia2010}
{Silvia}, D.~W., {Smith}, B.~D., \& {Shull}, J.~M. 2010, \apj, 715, 1575

\bibitem[{{Sonneborn} {et~al.}(1998){Sonneborn}, {Pun}, {Kimble}, {Gull},
  {Lundqvist}, {McCray}, {Plait}, {Boggess}, {Bowers}, {Danks}, {Grady},
  {Heap}, {Kraemer}, {Lindler}, {Loiacono}, {Maran}, {Moos}, \&
  {Woodgate}}]{Sonneborn1998}
{Sonneborn}, G., {et~al.} 1998, \apjl, 492, L139

\bibitem[{{Staveley-Smith} {et~al.}(1993){Staveley-Smith}, {Briggs}, {Rowe},
  {Manchester}, {Reynolds}, {Tzioumis}, \& {Kesteven}}]{Staveley-Smith1993}
{Staveley-Smith}, L., {Briggs}, D.~S., {Rowe}, A.~C.~H., {Manchester}, R.~N.,
  {Reynolds}, J.~E., {Tzioumis}, A.~K., \& {Kesteven}, M.~J. 1993, \nat, 366,
  136

\bibitem[{{Sugerman} {et~al.}(2005){Sugerman}, {Crotts}, {Kunkel}, {Heathcote},
  \& {Lawrence}}]{Sugerman2005}
{Sugerman}, B.~E.~K., {Crotts}, A.~P.~S., {Kunkel}, W.~E., {Heathcote}, S.~R.,
  \& {Lawrence}, S.~S. 2005, \apj, 627, 888

\end{thebibliography}
\end{document}